\documentstyle[aps,twocolumn,floats]{revtex}
\input psfig.sty

\begin{document}
\draft
\title{Low Energy Excitations of Yb$_4$As$_3$ in a Magnetic Field}
\author{Yuri Kudasov$^{*,\dag}$, Gennadi Uimin$^{*,\S}$, Peter Fulde$^*$, Alexander
Ovchinnikov$^*$}
\address{$^*$Max-Planck-Institut f\"ur Physik komplexer Systeme, 
Dresden, Germany}
\address{$^\dag$Russian Federal Nuclear Center - VNNIEF, Sarov, Russia}
\address{$^\S$Landau Institute for Theoretical Physics, Moscow, Russia}
\date{August 25, 1999}
\maketitle

\begin{abstract}
We discuss the effects of an applied magnetic field on the low energy
excitations in the low temperature phase of Yb$_4$As$_3$. We show also why
the magnetic interaction of the Yb$^{3+}$ ions is nearly of an isotropic
Heisenberg spin-1/2 type. A small anisotropy due to an intrachain dipolar
interaction leads to the opening of a gap when a magnetic field is applied.
The model agrees with available experimental data. Simple experiments are
suggested in order to further test the present theory.
\end{abstract}

\pacs{71.27.+a,  75.30.Ds, 75.20.Hr}

At low temperatures the rare-earth pnictide Yb$_4$As$_3$ is a semimetal with
characteristic features of a heavy fermion system \cite{kasuya1,kasuya2}.
The linear specific heat coefficient is large, $\gamma \approx 
$ 200 mJ/mol K$^2$, and the spin susceptibility is enhanced accordingly. 
The resistivity
is of the Fermi-liquid type, i.e., $\rho(T) = \rho_0 + AT^2$ and the ratio 
$A/\gamma^2$ is of the same order of magnitude as for other heavy fermion
systems. Another feature of the low-energy phase is a low carrier
concentration. Measurements of the Hall coefficient yield approximately one
charge carrier per $10^3$ Yb ions. When a magnetic field is applied a gap in
the excitation spectrum seems to open up. This refers to measurements of the
specific heat \cite{helfrich1,helfrich2} where it has been found that a
field of 4 T leads to a dramatic decrease of the linear term in the specific
heat below 0.5 K. Two different proposals have been made to account for this
effect \cite{fulde2,oshikawa}. They are described below. Here we introduce
an alternative explanation and discuss a simple experiment which should be
able to discriminate between the different underlying physical pictures.

In order to understand the physical issue we have to recall some of the
basic properties of Yb$_4$As$_3$. At high temperatures the compound has the
anti-Th$_3$P$_4$ structure which is of cubic symmetry (T$_d^6$). The Yb ions
(Yb$^{+3}$:Yb$^{+2}$=1:3) occupy four families of interpenetrating chains
oriented along the diagonals of a cube. At room temperature a phase
transition to a trigonal low-temperature phase takes place. It is
accompanied by a charge ordering of the Yb$^{3+}$ 4$f$ holes 
\cite{helfrich2}. As the temperature decreases they align in the chains along the trigonal
direction, e.g., $\langle 111 \rangle$ and the system becomes a semimetal 
\cite{fulde2}. Since the 4$f$ holes are strongly correlated the system
behaves at low temperatures like one of well separated spin chains. Indeed,
inelastic neutron scattering (INS) experiments by Iwasa {\it {et al.}} 
\cite{iwasa} and Kohgi ${\it {et}}$ ${\it {al.}}$ \cite{koghi,mignot}
have demonstrated that the magnetic excitations are well described by means
of a one-dimensional isotropic Heisenberg chain, i.e., by de Cloiseaux -
Pearson spectrum \cite{dCP}, $\epsilon(q) = \frac 12 \pi J_{\rm eff} |\sin q|,
 -\pi \leq q \leq \pi$. They found $J_{{\rm eff}}$ to be $\sim 25$ K.
Since no magnetic ordering was observed down to 0.045 K \cite{bonville} the
interchain coupling must be very weak \cite{fulde2}.

The present work aims to shed light on two main problems: why is the
interaction of the ordered Yb$^{3+}$ ions in the chains so well described by
an {\sl isotropic} Heisenberg Hamiltonian, i.e., without a sizeable
anisotropy? What is the effect of an applied magnetic field on the
low-energy excitation spectrum? As regard the last topic, two models have
been put forward. One is based on intrachain interactions. By assuming a
ratio of $J^{\prime}/J_{{\rm eff}} \approx 10^{-4}$, for the inter-($%
J^{\prime}$) to intra ($J_{{\rm eff}}$) chain coupling constants, the low
temperature specific heat in a magnetic field can be well described. The
other model is due to Oshikawa ${\it {et}}$ ${\it {al.}}$ \cite{oshikawa}
and links the opening of a gap to an effective staggered field introduced by
an alternating $g$-tensor and Dzyaloshinsky-Moriya interaction. With this
interesting model predictions are made for the dependence of the gap in the
excitation spectrum on the direction of the magnetic field. The mechanism we
want to suggest here is quite different from the previous cases. A simple
specific heat experiment should be able to discriminate between the
different suggestions.

In order to derive the magnetic interactions in the Yb$^{3+}$ chains we
start from the Hamiltonian of the 4$f$ holes. In general, the intra-$f$ band
hopping amplitude is not diagonal with respect to the angular momentum
projections $m$. However, if the global quantization axis is chosen parallel
to the chain axis, the hopping amplitude becomes diagonal in $m$. Indeed, in
this case the angular parts (spherical harmonics $(Y_3^m)^*$ and 
$Y_3^{m^{\prime}}$) of the wave functions located on sites $i$ and $j$ depend
on the {\it common} polar angle $\phi$. Thus, integration of the factor 
$\,\exp i(m^{\prime}-m)\phi\,$ results in Kronecker's 
$\delta_{m,m^{\prime}}$. 
The choice of the quantization axis parallel to all Yb$^{3+}$ chains
significantly simplifies the initial Hamiltonian: 
\begin{eqnarray}  \label{chainHam_diag}
{\cal H} = - \sum_{\langle ij \rangle} \sum_{m=-L}^{L} \sum_{\sigma = \pm
1/2} t(m) \,f^\dag_{im\sigma}f_{jm\sigma} + {\cal H}_{{\rm corr}},
\end{eqnarray}
where $t(m) = t(-m)$ and $\langle ij \rangle$ denotes pairs of nearest
neighbors in the chain. The other relevant interactions are contained in 
${\cal H}_{{\rm corr}}$. They consist of generalizations of the Hubbard-$U$
term, describing the ionic charge excitations. They also contain the on-site
spin-orbit coupling in accordance with the Russel-Saunders coupling scheme. 
Wave functions of $f$-electrons are extremely anisotropic and therefore $t$
depends strongly on $m$. In \cite{fulde1}, characteristic values of $t$'s
have been approximately estimated as 50 meV, while a typical Coulomb-like $U$
term is of order $\sim 10$ eV. 
The spin-orbit coupling is considerable weaker than the Coulomb energy and
is of order eV.

The interaction between localized 4$f$ holes is derived by second-order
perturbations theory. In doing so we have to project all $L = 3$, $S = 1/2$
states onto the lowest $J$ multiplet which for Yb$^{3+}$ is $J = 7/2$. After
a straight forward calculation we find for the leading term of the
interaction

\begin{eqnarray}  \label{RKKY}
{\cal H}_{{\rm m}} = \sum_{\langle ij \rangle} \sum_{\mu, \nu= - J}^J
T_{ij}(\mu,\mu ^{\prime}) f^\dag_{i\mu}f_{i\nu}f^\dag_{j\nu}f_{j\mu},
\end{eqnarray}

\noindent where $T(\mu, \nu) = T(\nu, \mu) \equiv T(| \mu |, | \nu |)$ and 
$\mu, \nu$ are projections of $J$ onto the chain axis. If the axis of
quantization does not coincide with the chain direction the form of 
${\cal H}_{\rm magn}$ is more complicated. Next we want to show that for 
Yb$^{3+}$
ions (\ref{RKKY}) reduces to an isotropic Heisenberg Hamiltonian. For this
purpose the crystalline electric field (CEF) has to be taken into account.
In trigonal symmetry the $^2F_{7/2}$ multiplet splits into four doublets.
From INS experiments \cite{koghi} the excitation energies from the ground
state (GS) doublet are known to be 14, 21 and 29 meV. Since those energies
are much larger than $J_{{\rm eff}}$ we have to project 
${\cal H}_{\rm magn}$ onto the GS doublet. In order to find the corresponding GS wavefunctions
we use the CEF Hamiltonian in $C^6_{3v}$ symmetry. For such a symmetry the
following forms are valid for the four doublets: 
$\alpha_i |\pm 7/2\rangle +
\beta_i |\pm 1/2\rangle + \delta_i |\mp 5/2\rangle, \;(i = 1,2,3)\,$ and 
$\,|\pm 3/2\rangle$. We exclude the last doublet for being the GS and make 
for the GS doublet $|+\rangle$, $|-\rangle$ the following ansatz

\begin{eqnarray}  \label{GS}
|\pm\rangle = \alpha |\pm 7/2\rangle + \beta |\pm 1/2\rangle + \delta |\mp
5/2\rangle.
\end{eqnarray}

In the next steps the matrix elements of ${\cal H}_{\rm magn}$ with
respect to the GS doublets of neighboring sites labeled 1 and 2, i.e., 
$|+_1, +_2\rangle, |+_1, -_2\rangle, |-_1, +_2\rangle,$ and 
$|-_1, -_2\rangle$
are determined. The only non-vanishing ones are 
$\langle {\pm}_1, {\pm}_2|{\cal H}_{\rm magn}|{\pm}_1, {\pm}_2\rangle$
and $\langle \pm_1, \mp_2|{\cal H}_{{\rm magn}}|\mp_1,
\pm_2\rangle$. From ({\ref{GS}) it follows that those matrix elements are
all equal, implying that we deal with an ideal Hamiltonian of state
permutations. We denote this value of the matrix element by $J_{{\rm eff}}/2$
and introduce pseudo-spin operators $\tau^\pm_i, \tau^z_i$ which act on the
GS doublet as follows, $\tau^\pm | \mp \rangle = | \pm \rangle$, $\tau^z |
\pm \rangle = \pm \frac{1}{2} | \pm \rangle$. The effective magnetic
exchange Hamiltonian is then of the form

\begin{eqnarray}  \label{5}
{\cal H}_{{\rm eff}} = J_{{\rm eff}} \sum_{\langle ij \rangle} 
\left(\bbox\tau_i\bbox\tau_j + \frac{1}{4}\right).
\end{eqnarray}

\noindent Its gapless spectrum leads to the observed low temperature
specific heat and the observed large $\gamma$ value can be well explained by
the measured size of $J_{\rm eff}$.

Let us now express the Zeeman energy 
${\cal H}_{Ze} = -g\mu_B {\bf H}\cdot {\bf J}$ 
in terms of the pseudo-spin ${\bbox\tau}$. A straight forward
calculation yields the following matrix elements:

\begin{eqnarray}  \label{matr_elements_zeeman}
\begin{array}{l}
\langle \pm|J_z|\pm\rangle = \pm \frac 12\, (7\alpha^2 +\beta^2 - 5\delta^2)
= \pm \frac 12\, j_{1}, \\ 
\langle \pm|J_x|\mp\rangle = \pm i \,\langle \pm|J_y|\mp\rangle = 
\sqrt{7}\alpha\delta + 2\beta^2 = \frac 12\, j_{2}.
\end{array}
\end{eqnarray}

\noindent The Zeeman term can therefore be written in the compact form 
${\cal H}_{{\rm Ze}}=-g\mu _{B}\sum_{i}(j_{1}H_{z}\tau_{i}^{z}+
j_{2}(H_{x}\tau _{i}^{x}+H_{y}\tau _{i}^{y}))$ which clearly
demonstrates that the effect of the magnetic field depends on its direction
relative to that of the chains. Despite the magnetic field anisotropy the
spectrum remains gapless provided the Zeeman energy remains less than 
$J_{\rm eff}$ so that a transition to a ferromagnetic state can be excluded.
The Bethe ansatz solution shows that the excitation energy goes to zero at a
wave vector $q_{H}$ which depends on ${\bf H}$ and shifts continously from 0
to $\pi $ as the field is increased (see for example {\cite{shiba}}).

A gap in the excitation spectrum opens up though, when the weak magnetic
dipolar interaction within a chain is taken into account. It is of the form

\begin{eqnarray}  \label{7}
{\cal H}_{dip} = g^2\mu_B^2 \sum_{i<j} 
\frac{{\bf J}_i\cdot {\bf J}_j - 
3({\bf J}_i\cdot{\bf n}) ({\bf J}_j\cdot{\bf n})}
{|{\bf R}_i - {\bf R}_j|^3} ,
\end{eqnarray}

\noindent where ${\bf n} = ({\bf R}_i - {\bf R}_j)/|{\bf R}_i - {\bf R}_j|$
and the ${\bf R}_i$ denote the positions of the Yb$^{3+}$ ions. We compute
the non vanishing matrix elements as enlisted

\begin{eqnarray}  \label{m_dipolar}
\begin{array}{l}
\langle \pm_1, \pm_2| J^z_1 J^z_2|\pm_1, \pm_2\rangle = 
\frac 14 j_1^2, \\ 
\langle \mp_1, \pm_2| J^z_1 J^z_2|\mp_1, \pm_2\rangle = 
- \frac 14 j_1^2, \\ 
\langle \pm_1, \mp_2| (J^+_1 J^-_2+J^-_1 J^+_2)|\mp_1,\pm_2\rangle = 
j_2^2 .
\end{array}
\end{eqnarray}

\noindent With their help we can write the interaction projected onto the
respective GS doublets as follows:
$ g^2\mu_B^2\sum_{i<j}(-2 j_1^2 \tau_1^z \tau_2^z + 
j_2^2 (\tau_1^x \tau_2^x + \tau_1^y \tau_2^y))/a^3$ 
where $a$ is the distance between neighboring Yb$^{3+}$ ions. 
This results in the following interaction Hamiltonian of the
Yb$^{3+}$ ions in the chain

\begin{eqnarray}  \label{total_ham}
{\cal H} = \sum_{\langle ij\rangle}\{(1 - \lambda_1) \tau^z_i \tau^z_j 
+ (1+\lambda_2) (\tau^x_i \tau^x_j + \tau^y_i \tau^y_j)  \nonumber \\
- \sum_i (h_x \tau^x_i + h_y \tau^y_i + h_z\tau^z_i)
\end{eqnarray}

\noindent with $\lambda_1 = 2 g^2\mu_B^2j_1^2\,/(J_{\rm eff}a^3)$, 
$\lambda_2 = g^2\mu_B^2 j_2^2\,/(J_{{\rm eff}}a^3)$, 
$h_z = g\mu_BH_zj_1/J_{{\rm eff}}$, 
$h_x = g\mu_BH_xj_2/J_{{\rm eff}}$, and 
$h_y = g\mu_BH_yj_2/J_{{\rm eff}}$. 
Until now we have not discussed possible RKKY type of
interactions between the Yb$^{3+}$ ions. 
Since the carrier concentration is
so low and since the specific heat behaves similar in the insulator 
Yb$_4$(As$_{0.6}$P$_{0.4}$)$_3$ as it does the semimetal Yb$_4$As$_3$ 
they may be safely neglected.

In order to determine $j_{1}$ and $j_{2}$ one can use the work of Griffiths 
\cite{griffith} on the 1D antiferromagnetic Heisenberg model. It allows for
deducing the values of $j_{1}$ and $j_{2}$ from longitudinal and transversal
magnetization data, ($M_{z},H_{z}$) and ($M_{x},H_{x}$), respectively. 
Unfortunately, monocrystal measurements
have been done for one direction only \cite{koghi}, that is insufficient to
determine both $j_{1}$ and $j_{2}$. However, by using a point-charge model
for the CEF set up by the octahedral ligands one can estimate that 
$j_{1}\approx j_{2}\approx 3.2$. If we assume that $j_{1}$ and $j_{2}$ are
nearly equal we obtain from the magnetization measurements $j_{1}\approx
j_{2}\approx 3$ in an agreement with the point-charge model calculation.
Then, the anisotropy parameters in (\ref{total_ham}) are 
$\lambda_{1}\approx 2\lambda _{2}\approx 10^{-2}$. 
The small anisotropy is in an
agreement with INS experiments \cite{mignot}. 

The excitation spectrum of the effective Hamiltonian ({\ref{total_ham}}) is
investigated by means of the spin-wave theory. Assuming a N\'{e}el state for
the chain and a small anisotropy we obtain (units of $J_{{\rm eff}}$): 
\begin{eqnarray}
\epsilon ^{2}(k) =x^{2}(1+Gc^{2})+y^{2}(1-Gc^{2})+z^{2}(1+c^{2})\pm 
\nonumber \\
c\sqrt{(x^{2}(1+G)+y^{2}(1-G)+2z^{2})^{2}-4x^{2}y^{2}(1-G^{2})}
\label{spectrum}
\end{eqnarray}
where $z=h_{z}/(1+G),x=h_{x}/2,y=1-x^{2}-z^{2},$ $G=1-2\lambda ,$ $\lambda
=(\lambda _{1}+\lambda _{2})/2,$ and $c=\cos k.$ The dipolar interaction
splits the magnetic excitation spectrum of the chain into two branches. But
in zero field the spectrum remains gapless. The same holds true when the
magnetic field is along the chain axis. But a field component perpendicular
to the chain, i.e., in the easy plane induces a gap in the spectrum. The two
branches yield the following gaps ({\ref{spectrum}}): $\Delta _{1}=h_{x}$
and $\Delta _{2}=2\sqrt{\lambda (1-h_{x}^{2}/4)}$. Therefore we expect that
the specific heat of a single Yb$_{4}$As$_{3}$ crystal behaves very
anisotropic in an applied field. This suggests the following scenario for
the low temperature specific heat $C(T)$; (i) at small field $h_{x}$ it is 
$\Delta _{1}<\Delta _{2}$ and the smaller gap is linear in the magnetic
field; (ii) as the field increases $\Delta _{1}$ becomes eventually larger
than $\Delta _{2}$. This should be the case for fields between 2T and 4T
when the above parameter values are used. At higher fields the specific heat
should change little when temperature is smaller than $\Delta _{2}$ which
amounts to about 2K, because $\Delta _{2}$ changes only slowly with $h_{x}$.
There is some reduction in $\gamma$ at higher temperatures since 
$\Delta _{1}$ continues to increase linearly in $h_{x}$; 
(iii) at very high fields $\Delta_{2}$ starts decreasing and eventually 
goes to zero at $h_{x}=2$ (about 25T).
This is practically the same field at which the transition to the
ferromagnetic state should take place. Since the spectrum is almost
quadratic in the vicinity of $q=0$ and $q=\pi $ at $h_{x}=2$, $\gamma$
starts growing $\propto T^{-1/2}$, reaching $\sim \lambda ^{-1/2}$ at a
its maximum.

In order to make detailed predictions we have calculated numerically $\gamma$
 for polycrystalline Yb$_{4}$As$_{3}$ by taking an average over all
directions. Thereby the features discussed above are smoothened out a bit.
In Fig.\ 1 we show the results for $\gamma $ when the magnetic field is
relatively small. The region between curves $a$ and $b$ corresponds to the case 
$\Delta _{1}<\Delta _{2}$. The results are in satisfactory agreement with the
experimental data \cite{helfrich2}. Curves $c$ and $d$ are very close to each
other at low $T$ since the cross-over to $\Delta _{1}>\Delta _{2}$ is taking
place. A hint of that cross-over was recently observed in \cite{Koppen}
where a saturation of the gap at fields higher than 4T was detected by
specific heat measurements. The behavior of the latter at high fields is
shown in Fig.\ 2.

\begin{figure}[h]
\centerline{\psfig{file=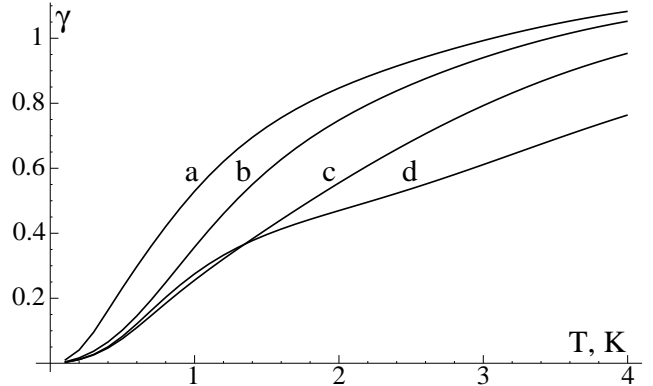,width=8.5cm}}
\caption[l]{The linear term of the specific heat for polycrystalline Yb$_4$As$_3$
calculated from spin excitations (\ref{spectrum}). The low-temperature value of $C/T$ 
at h=0 is taken as an unit.
$J_{\rm eff}=25$ K, $\lambda_1=2\lambda_2=0.01$, 
$j_1=j_2=3$. $B=1$T (a), $B=2$T (b), $B=4$T (c), 
and $B=8$T (d).}
\label{fig1}
\end{figure}

\begin{figure}[h]
\centerline{\psfig{file=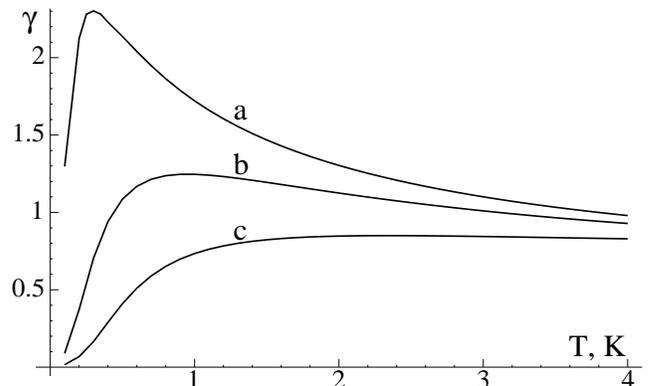,width=8.5cm}}
\caption[l]{The linear term of the specific heat for polycrystalline 
Yb$_4$As$_3$ at high magnetic field. (a) $h=2$, (b) $h=1.9$, 
and (c) $h=1.7$, other parameters are
the same as in Fig.\ref{fig1}.}
\label{fig2}
\end{figure}
We suggest therefore specific heat measurements on Yb$_{4}$As$_{3}$ in
fields up to 30T. They should be able to confirm or refute the explanation
suggested here for the opening of a gap in the excitation spectrum by an
applied field. Such measurements should first show a depletion of the low
energy excitation as the field strength increases. At high fields the gap
should close again and the specific heat at low temperatures should become
even larger than in the absence of a field.


We would like to thank K. Ueda for discussions and providing us with a
preprint of his work prior to publication.


\end{document}